\documentclass[aps,prb,superscriptaddress,notitlepage,nopacs,amsmath,amstex,amssymb,citeautoscript,longbibliography,floatfix,letter,twocolumn,10pt]{revtex4-2}
\usepackage{graphicx}
\usepackage{subfigure}
\usepackage{adjustbox}
\usepackage{bm}
\usepackage{color}
\usepackage{braket}
\usepackage{standalone}
\usepackage{multirow}
\usepackage{tikz}
\usepackage{mathrsfs}
\usepackage{dsfont}
\usepackage{comment}
\usepackage{amsmath}
\usepackage[colorlinks,bookmarks=true,citecolor=blue,linkcolor=red,urlcolor=blue]{hyperref}
\usepackage{cleveref}
\usepackage{orcidlink}
\renewcommand{\vec}[1]{\mbox{\boldmath$#1$}}

\begin{document}

\title{
Entanglement scaling and charge fluctuations in a Fermi liquid of composite fermions
} 

\author{Cristian Voinea\orcidlink{0000-0002-8086-3791}}
\affiliation{School of Physics and Astronomy, University of Leeds, Leeds LS2 9JT, United Kingdom}

\author{Songyang Pu\orcidlink{0000-0003-0144-1548}}
\affiliation{School of Physics and Astronomy, University of Leeds, Leeds LS2 9JT, United Kingdom}
\affiliation{Department of Physics and Astronomy, The University of Tennessee, Knoxville, TN 37996, USA}

\author{Ajit C. Balram\orcidlink{0000-0002-8087-6015}}
\affiliation{Institute of Mathematical Sciences, CIT Campus, Chennai 600113, India}
\affiliation{Homi Bhabha National Institute, Training School Complex, Anushaktinagar, Mumbai 400094, India}

\author{Zlatko Papi\'c\orcidlink{0000-0002-8451-2235}}
\affiliation{School of Physics and Astronomy, University of Leeds, Leeds LS2 9JT, United Kingdom}

\date{\today}

\begin{abstract}
The composite fermion Fermi liquid (CFL) state at $\nu=1/2$ filling of a Landau level is a paradigmatic non-Fermi liquid borne out purely by Coulomb interactions. But in what ways is this exotic state of matter different from a Fermi liquid? The CFL entanglement entropy was indeed found to exhibit a significant enhancement compared to free electrons [Shao \emph{et al.}, Phys. Rev. Lett. {\bf 114}, 206402 (2015)], which was subsequently ruled out as a finite-size effect by the study of a lattice CFL analog [Mishmash and Motrunich, Phys. Rev. B {\bf 94}, 081110 (2016)]. Moreover, the enhancement was not observed in a quasi-one-dimensional limit of the Coulomb ground state at $\nu=1/2$ [Geraedts \emph{et al.}, Science {\bf 352}, 197 (2016)].
Here, we revisit the problem of entanglement scaling in the CFL state realized in a two-dimensional electron gas. Using Monte Carlo evaluation of the second R\'enyi entropy $S_2$ for the CFL variational wave function, we show that the entanglement enhancement is present not only at $\nu=1/2$ but also at $\nu=1/4$, as well as in bosonic CFL states at $\nu=1$ and $\nu=1/3$ fillings. In all cases, we find the scaling of $S_2$ with subsystem size to be enhanced compared to the non-interacting case, and insensitive to the choice of geometry and projection to the lowest Landau level. 
We also demonstrate that, for CFL states, the variance of the particle number in a subsystem obeys area-law scaling with a universal subleading corner contribution, in stark contrast with free fermions. Our results establish the enhanced entanglement scaling and suppressed charge fluctuations as fingerprints of non-Fermi-liquid correlations in CFL states.
\end{abstract}

\maketitle

\section{Introduction}

The fundamental quasiparticles of fractional quantum Hall states are composite fermions (CFs) -- electrons dressed by an even number of vortices~\cite{Jain89}. At even-denominator fillings of the lowest Landau level (LLL), the CFs can form a compressible state -- the ``composite fermion Fermi liquid'' (CFL)~\cite{HalperinLeeRead, Pasquier98, Read98}, which has been observed in numerous experiments~\cite{Shayegan20b}. The CFL state has also been argued to emerge in systems with flat bands in the absence of a magnetic field, such as twisted moir\'e materials \cite{Dong23, Goldman23}. 

Flux attachment -- the mechanism behind the formation of CFs~\cite{Lopez91, HalperinLeeRead, Jain07} -- couples the Fermi surface of CFs with an internal statistical U(1) gauge field, placing the CFLs inside a larger class of ``non-Fermi liquids" (NFLs) (see~Ref.~\cite{ChowdhuryRMP} for a recent review).
On the one hand, the CFL is a pristine example of a stable NFL phase that emerges solely due to electron-electron interactions. On the other hand, the existence of well-defined CF quasiparticles implies there may be a qualitative similarity between CFL and conventional Fermi liquids~\cite{Son15, Wang16, Halperin20b}. Indeed, the Fermi wave vector of CFs is consistent with that derived from the electron density~\cite{Kamburov14b, Kamburov14c, Mueed15} and it also satisfies the Luttinger theorem~\cite{Balram15c, Balram17}. Thus, pinpointing the NFL nature of the CFL is a nontrivial task. 

One sensitive diagnostic of a quantum state is its bipartite entanglement entropy. The entanglement entropy scaling in non-interacting Fermi liquids obeys the well-known Widom formula \cite{widom1982, Gioev06, Leschke14}, wherein the ``area law'' -- ubiquitous to gapped systems -- gets modified by a multiplicative logarithmic correction. It has been argued that turning on interactions does not modify the scaling \cite{SwingleSenthil2013}, leaving open the value of the prefactor~\cite{McMinis2013, Hu20}. Previous numerical tests of this hypothesis for the CFL state have arrived at conflicting conclusions. 
Infinite density matrix renormalization group (iDMRG) study of the Coulomb ground state at $\nu=1/2$ revealed no sign of correction to the Widom formula~\cite{Geraedts16}. 
For the continuum CFL variational wave function of Rezayi and Read~\cite{Rezayi94}, Ref.~\cite{Shao15} found a significant multiplicative prefactor to the entanglement entropy scaling ($\approx2$ in system sizes on the order of $40$ electrons). Lastly, Ref.~\cite{Mishmash16} also found an enhancement for a lattice analog of the CFL wave function. However, based on the scaling of different contributions to the entanglement from the sign and modulus of the wave function, Ref.~\cite{Mishmash16} ruled out the enhancement as a finite-size effect. These studies therefore raise an important question: is there any distinctive entanglement signature of the CFL state in the thermodynamic limit?

Here we report a systematic study of entanglement in several CFLs described by the Rezayi-Read wave function~\cite{Rezayi94}, known to have a large overlap with the LLL Coulomb ground state~\cite{Rezayi00, Liu20}. We assume full spin polarization and compare the fermionic CFLs at fillings $\nu=1/2$ and $\nu=1/4$, which are distinguished by the number of vortices attached to each electron (two and four, respectively), and contrast those against an odd number of vortices in bosonic CFLs at $\nu=1$ and $\nu=1/3$ fillings. 
We employ Monte Carlo techniques to evaluate the second R\'enyi entropy using the SWAP algorithm \cite{Hastings10}. In all CFL states considered, we find a pronounced enhancement of entanglement scaling, consistent with Ref.~\cite{Shao15}, up to the largest system sizes accessible ($N\lesssim 60$). At the same time, we reconcile these findings with entropy decompositions considered in Ref.~\cite{Mishmash16} by showing that the latter is more sensitive to finite-size effects and LLL projection compared to the total entropy. We demonstrate the robustness of our conclusions in sphere and torus geometries. Finally,  we use the same setup to evaluate charge fluctuations in a subsystem, for which analytical predictions are available~\cite{Wu2024bipartite}. We confirm that charge fluctuations in the CFL state follow the expected area-law scaling, in sharp contrast with free fermions. Moreover, within the same range of system sizes, we accurately extract the subleading corner contribution to the charge fluctuations, in agreement with the expected ``super-universal'' value~\cite{Wu2024bipartite} that also holds for gapped~\cite{Estienne22, Berthiere2023} and critical states~\cite{Wu2021, Wang2021} in (2+1)-dimension. 

The remainder of this paper is organized as follows. In Sec.~\ref{sec:CFL} we introduce the CFL wave functions on the torus and sphere geometries and evaluate their entanglement entropy. Section~\ref{sec:charge} contains the results for charge fluctuations in the same CFL states. In Sec.~\ref{sec:decomp} we discuss the scaling of entanglement entropy in the context of modulus and sign entropy decompositions, carefully contrasting the results against free fermions. Our conclusions and a discussion of open questions are presented in Sec.~\ref{sec:conclusion}, while Appendices contain further technical details, including a description of the Monte Carlo algorithm and the LLL projection of the wave functions.

\section{The CFL wave function and its R\'enyi entropy}\label{sec:CFL}

The essential physics of the CFL state at filling factor $\nu=1/m$ is captured by the ansatz wave function~\cite{Rezayi94,HalperinLeeRead}:
\begin{eqnarray}\label{eq:WF}
\Psi_{m}^\mathrm{CFL}(z_1,\ldots,z_N) = \mathrm{Det} \big[ 
\chi_{n} (z_{j},z^{*}_{j}) \big] \psi_m^\mathrm{L}(z_1,\ldots,z_N),
\end{eqnarray}
expressed in terms of particle coordinates $z_j = x_j + i y_j$.
The first term in Eq.~\eqref{eq:WF} is a Slater determinant of single-particle orbitals $\chi_{n}$ occupied by CFs in zero magnetic field, while the second  is the Laughlin-Jastrow wave function at filling factor $\nu=1/m$~\cite{Laughlin83}. Due to the antisymmetry of the Slater determinant, these wave functions describe fermionic CFL states for $m$ even and bosonic CFL states for $m$ odd. As discussed below, the wave functions of Eq.~\eqref{eq:WF} can be adapted to the spherical and toroidal geometries considered in this work, see Refs.~\cite{Rezayi94, Fremling18, Geraedts18}. Moreover, the wave function of Eq.~\eqref{eq:WF} is not constrained to reside in the LLL. Therefore, to make contact with previous work, we will project the wave function to the LLL using the Jain-Kamilla (JK) method~\cite{Jain97, Shao15, Pu17, Pu18}. 
Nevertheless, we will probe the sensitivity of our results by also studying the unprojected wave function as in Eq.~\eqref{eq:WF}.  

\begin{figure*}[htb]
    \centering \includegraphics[width=0.95\linewidth]{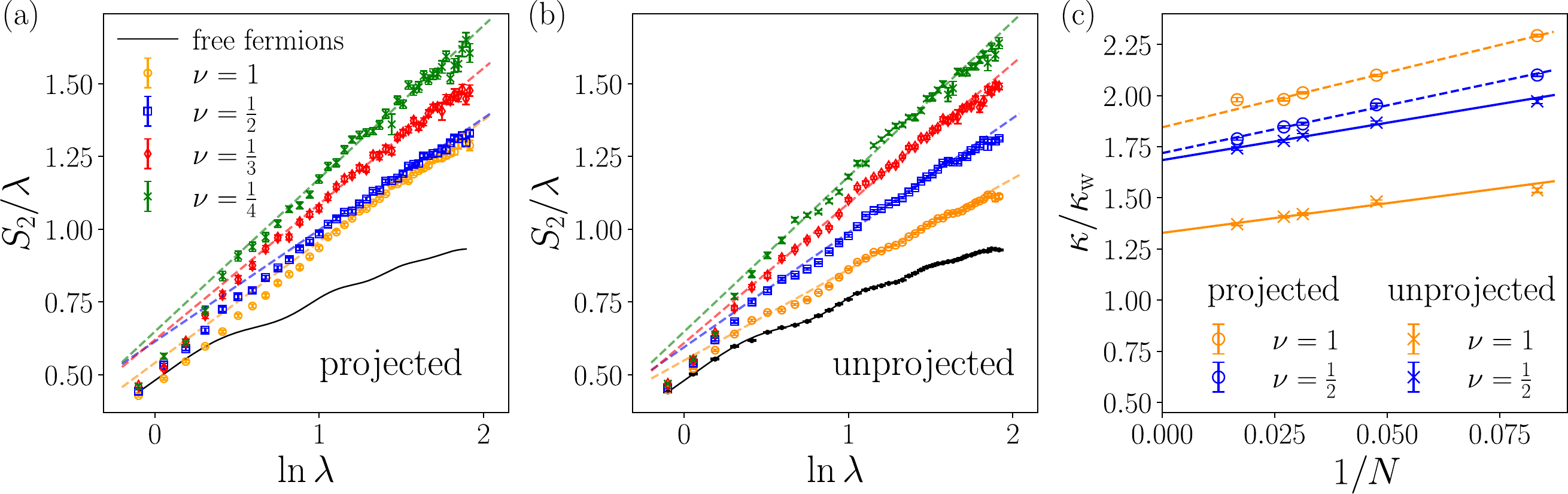}
    \caption{(a)-(b): $S_2$ R\'enyi entropy for projected (a) and unprojected (b) CFL wave functions at fillings $\nu  = 1/m$, with $m=1,2,3,4$. For all fillings, $S_2$ exhibits a multiplicative correction $\sim1.7$ to the Widom formula, Eq.~\eqref{eq: Widom_formula}. The $2k_F$ Friedel oscillations~\cite{HalperinLeeRead} are visible in all cases. All data is for $N{=}37$ particles on a square torus. The symbols represent Monte Carlo data, while the continuous black line is the exact diagonalization result for free fermions. (c) Extrapolations of the ratio between the slope $\kappa$ ($S_2$ vs. $\lambda \ln \lambda$) of the CFL at fillings $\nu=1/2$ and $\nu=1$ and the free-fermion slope $\kappa_{\mathrm{w}}$. We used system sizes $N=12-60$ and data points in the range which yield the extrapolated free-fermion slope $\approx0.25$, see Appendix~\ref{app:extrapolation} for details.
    } 
    \label{fig: torus total entropy}
\end{figure*}

The central object of our study is the second R\'enyi entropy of the reduced density matrix $\hat\rho_A$ of region $A$:
\begin{eqnarray}
S_2 = - \ln \left( \operatorname{tr}_A \hat \rho_A^2 \right), \quad  \hat \rho_A {=} \operatorname{tr}_{\bar A}|\Psi_m^\mathrm{CFL}\rangle\langle\Psi_m^\mathrm{CFL}|,
\end{eqnarray}
obtained by tracing over the complement $\bar A$ of region $A$. 
We typically take $A$ to be a circle of radius $r_{A}$, although in Sec.~\ref{sec:charge} we will also consider a square-shaped region to determine corner contributions to charge fluctuations in $A$. To extract the scaling, we fix a large system size and vary $r_{A}$. $S_2$ can be conveniently evaluated via 
variational Monte Carlo by replacing the trace with an expectation value of a SWAP operator between two copies of the system~\cite{Hastings10}, which is reviewed in Appendix~\ref{app: algorithm and decomp}.

For any wave function $\Psi$ that can be written as a Slater determinant, $\Psi={\rm Det}[\phi_m(\vec{r}_n)]$, where the single-particle orbitals $\phi_m$ are orthonormalized over the full space, the entanglement entropy can be efficiently computed using the correlation matrix $\mathcal{A}$~\cite{Chung01}. The matrix elements of $\mathcal{A}$ are given by the overlap integrals on the subsystem, $\mathcal{A}_{mn}=\int_A \mathrm{d}\mathbf{r}\,\phi_m^*(\mathbf{r})\phi_n(\mathbf{r})$. Denoting the eigenvalues of $\mathcal{A}$ by $a_m$, the second R\'enyi entropy is given by \cite{Klich06,Calabrese11}
\begin{equation}
\label{overlap}
    S_2^\mathrm{free}=-\sum_m \ln\left(a_m^2+(1-a_m)^2\right).
\end{equation}
Explicit expressions for $\mathcal{A}_{mn}$ in different geometries considered in this work are given in Appendix~\ref{app:overlaps}. With this, one can use Eq.~\eqref{overlap} to efficiently evaluate the R\'enyi entropy of free fermions in systems of finite size.

In the asymptotic limit, for free fermions in 2D, the area-law scaling of $S_2$ is violated by a term that depends log-linearly on the dimensionless quantity $\lambda \equiv k_{F} r_{A}$, with $k_{F}$ denoting the Fermi momentum. 
The violation is given by the Widom formula \cite{widom1982, Gioev06, Leschke14}:
\begin{equation} \label{eq: Widom_formula}
    S_{2}^\mathrm{free} = \kappa_{\mathrm{w}} \lambda \ln \lambda + \ldots, 
\end{equation}
where $\kappa_{\mathrm{w}}$ is the Widom coefficient that depends solely on the geometry and effective central charge of a (1+1)-dimensional chiral relativistic fermion~\cite{Swingle10, Swingle12}. 
For a circular 2D Fermi surface, we have $\kappa_{\mathrm{w}} = 1/4$. One of the goals of this paper is to ascertain whether the entropy of CFL states follows the scaling $S_2 = \kappa \lambda \ln \lambda+\ldots$, analogous to Eq.~\eqref{eq: Widom_formula} and whether $\kappa$ is the same as $\kappa_{\mathrm{w}}$. In the remainder of this section, we directly evaluate the $S_2$ R\'enyi entropy for the wave functions in Eq.~\eqref{eq:WF} on the torus and sphere, and compare them against Eq.~\eqref{eq: Widom_formula}.

\subsection{Torus}

On the torus, the CFL state at filling factor $\nu=1/m$ is described by the wave function in Eq.~\eqref{eq:WF}, where we take the single-particle orbitals to be plane waves, $\chi_{n} (\mathbf{r}_{j}) = \exp(i\mathbf{k}_{n}\cdot \mathbf{r}_{j})$, with $\mathbf{k}_{n}$ the wave vectors of the occupied states of the Fermi sea~\cite{Rezayi94}. 
The latter can be determined from an empirical rule given in Refs.~\cite{Geraedts18, Fremling18, Shao15}. On the other hand, the Laughlin-Jastrow part of the wave function $\psi^{\mathrm{L}}_{m}$ has a more complicated form given by 
\begin{eqnarray}  \label{eq:laughlintorus}
\psi^{\mathrm{L}}_{m} &=&   \vartheta \begin{bmatrix}
     \notag    \frac{N-1}{2} \\ \frac{m(N-1)}{2}
    \end{bmatrix} \bigg( \frac{m \sum z_{i}}{L_{1} } \bigg| m\tau\bigg) \\
  \notag  &&  \prod_{i<j} \bigg\{ \vartheta \begin{bmatrix}
        \frac{1}{2} \\ 
        \frac{1}{2}
    \end{bmatrix} \bigg( \frac{z_{i} - z_{j}}{L_{1}} \bigg| \tau\bigg) \bigg\}^m  \exp \bigg( {\sum_i \frac{z_{i}^{2} - |z_{i}^{2}|}{4 \ell^{2}}} \bigg),\\
    \end{eqnarray}
where the torus periodicity is ensured by the Jacobi $\vartheta$ functions with rational characteristics~\cite{Haldane85,Mumford07,Pu21}. The Jacobi $\vartheta$ functions explicitly depend on the modular parameter $\tau$ of the torus and one of its dimensions, $L_1$, where $\ell$ denotes the magnetic length. The $\vartheta$ function in the first line of Eq.~\eqref{eq:laughlintorus} describes the center of mass of the particles, and we have implicitly chosen it to carry zero momentum and obey strictly periodic conditions for the torus~\cite{Haldane85b}. The remaining terms in Eq.~\eqref{eq:laughlintorus} represent a periodized version of the familiar Laughlin-Jastrow factor in the infinite disk geometry~\cite{Laughlin83}. 

While the wave function $\psi_m^\mathrm{L}$ in Eq.~\eqref{eq:laughlintorus} is fully within the LLL, the multiplication via plane waves $\chi_{n} (\mathbf{r}_{j})$ brings the total wave function $\Psi^{\mathrm{CFL}}_{m}$ outside the LLL.  Following Ref.~\cite{Pu20b}, we obtain the family of LLL-projected CFL wave functions:
\begin{align}\label{eq:projtorus}
    \notag \Psi^{\mathrm{CFL,proj [\boldsymbol{\alpha}]}}_{m} &= \vartheta \begin{bmatrix}
       \frac{N-1}{2} \\ \frac{m(N-1)}{2}
    \end{bmatrix} \bigg( \frac{m \left(\sum z_{i} + i\ell^{2}\sum k_{i} \right)}{L_{1} } \bigg| m\tau\bigg) \\
     &\times \mathrm{Det} [g_{nl}^{[\boldsymbol{\alpha}]}] \exp \bigg( {\sum_i \frac{z_{i}^{2} - |z_{i}^{2}|}{4 \ell^{2}}} \bigg)\, , 
\end{align}
where the key consequence of LLL projection are the CF ``single-particle'' orbitals $g_{nl}^{\boldsymbol{\alpha}}$:
  \begin{align}\label{eq:g}
  \notag g_{nl}^{[\boldsymbol{\alpha}]} &= e^{k_{n}(k_{n}+2\overline{k}_{n})\ell^{2}/4} e^{i(k_{n}+\overline{k}_{n})z_{l}/2} \\
    &\times \prod_{p=1}^{m/2} \prod_{j\neq l} \vartheta \begin{bmatrix}
       \frac{1}{2} \\ \frac{1}{2}
    \end{bmatrix} \bigg( \frac{z_{l} + i\alpha_{p}k_{n}\ell^{2} - z_{j}}{L_{1} } \bigg| \tau\bigg),
\end{align}
where $\boldsymbol{\alpha} = (\alpha_{1},\dots,\alpha_{m/2})$ are the corresponding JK projection coefficients. Torus periodicity enforces the constraint $\sum_{p=1}^{m/2}\alpha_{p} = m$.

Note that  $g_{nl}^{\boldsymbol{\alpha}}$ in Eq.~\eqref{eq:g} depend on the coordinates of all the particles, hence they are complicated many-body functions that only formally resemble single-particle orbitals. Essentially, the LLL projection shifts the attached vortices by an amount proportional to the wave number of the CF. The scheme in Eq.~\eqref{eq:projtorus} encompasses fermionic CFL wave functions with even $m$; for bosons, one needs to attach one additional set of Jastrows outside the wave function for $m \geq 3$ and remove one for $m=1$. In Appendix~\ref{app:JK} we verify the consistency of our results for different choices of $\boldsymbol{\alpha} $, showing that, in general, our results are insensitive to the details of the projection.

In Fig.~\ref{fig: torus total entropy}(a)-(b) we show the evaluated  $S_2$ for bosonic and fermionic CFL states at fillings $\nu=1/m$, with $m {=} 1,2,3,4$. Based on our free-fermion benchmarks presented in Appendix~\ref{app:extrapolation}, to minimize the finite-size effect, we harvest data points such that the subsystem area ranges from 1\% to 30\% of the total area and use a fixed Fermi momentum of $k_{F} = \sqrt{2\nu}/\ell$. The extrapolated slope $\kappa$, in units of the Widom slope $\kappa_\mathrm{W}$, is plotted in Fig.~\ref{fig: torus total entropy}(c) as a function of inverse system size.

Our results for $\nu=1/2$ in Fig.~\ref{fig: torus total entropy}(a) are consistent with Ref.~\cite{Shao15} and scaling in Eq.~\eqref{eq: Widom_formula}, with the coefficient $\kappa \approx 1.7\kappa_{\mathrm{w}}$ significantly violating the Widom formula. By comparing the circular and square subregions, we find the violation to be independent of the subregion shape, see Appendix~\ref{app:extrapolation}. Another thing to note is that attaching a different number of vortices only has a weak effect on the slope $\kappa$, as seen from the parallel lines in Fig.~\ref{fig: torus total entropy}(a). Moreover, LLL projection also has an almost negligible effect on $\kappa$, except for the bosonic $\nu=1$ CFL state, as illustrated in Fig.~\ref{fig: torus total entropy}(c). The stronger effect of LLL projection on $\nu=1$ is expected based on the single vortex attached~\cite{SOM}. Next, we show that similar results are obtained for CFL states on the sphere.      

\subsection{Sphere}\label{sec:sphere}

The unprojected CFL wave function on the sphere is of the form in Eq.~\eqref{eq:WF}, where the single-particle orbitals are the standard spherical harmonics, $\chi_{n} (\mathbf{r}_{j}) = Y_{L_{n}M_{n}}(\theta_{j},\varphi_{j})$, where $L$, $M$ stand for the angular momentum and its $z$-component, respectively, while $\theta_j$ and $\phi_j$ are the polar and azimuthal angles describing the position of $j$th particle in terms of spherical angle $\mathbf{\Omega}_j$. The spherical Laughlin-Jastrow wave function has a compact expression
\begin{eqnarray}\label{eq:laughlinsphere}
 \psi^{\mathrm{L}}_{m} = \prod_{i<j} (u_{i}v_{j} - v_{i}u_{j})^{m}, 
\end{eqnarray}
in spherical spinor coordinates~\cite{Haldane83}, 
\begin{eqnarray}
    u_j = \cos (\theta_j/2) e^{i\phi_j/2}, \;\;\; v_j= \sin (\theta_j/2) e^{-i\phi_j/2}.
\end{eqnarray}
Note that the Widom formula for free fermions can also be appropriately modified for the sphere geometry, as shown in Appendix~\ref{app:widomsphere}.

Before presenting the results, we briefly explain how to perform the LLL projection on the sphere. Similar to the torus, while Eq.~\eqref{eq:laughlinsphere} describes a proper LLL state, the spherical harmonics have components in arbitrary LLs. Let us first treat the state at $\nu=1/2$, with two vortices attached i.e., $\Psi^{\rm CFL}_{1/2}{=}\mathcal{P}_{\rm LLL} \prod_{j{<}k} (u_{j}v_{k} {-} u_{k}v_{j})^2 \Phi_{\rm FS}$, where $\Phi_{\rm FS}$ is the filled-shell wave function on the sphere at zero field that is analogous to that of a Fermi sea~\cite{Rezayi94}. The projected state can be evaluated as a Slater determinant of CF wave functions $Y^{CF}_{0LM}$ which, similarly to the torus geometry, depend on the coordinates of all particles through the Jastrow factors $J_{j} {=} \prod_{k {\neq} j} (u_{j}v_{k} {-} u_{k}v_{j})$. We quote the final expression here: 
\begin{eqnarray}
    \notag Y^{CF}_{0LM} (\mathbf{\Omega}_{j}) &=& N_{0,L,M} (-1)^{L-M} \frac{(2Q+1)!}{(2Q+L+1)!} u_{j}^{M} v_{j}^{-M} \\
    \notag && \sum_{s=0}^{L} (-1)^{s} \begin{pmatrix} L \\ s
    \end{pmatrix} \begin{pmatrix} L \\ L-M-s
    \end{pmatrix} u_{j}^{s} v_{j}^{L-s} \\
    &&\bigg[ \bigg(\frac{\partial}{\partial u_{j}}\bigg)^{s} \bigg(\frac{\partial}{\partial v_{j}}\bigg)^{L-s}  J_{j} \bigg] \, ,
\end{eqnarray}
where $N_{0, L, M}$ is a normalization factor, and the flux for the physical electrons is $2Q=2(N{-}1)$. Given that we only have one power of Jastrow factors, the derivatives can be evaluated as follows~\cite{Jain07}:
\begin{eqnarray}
    \notag && \bigg(\frac{\partial}{\partial u_{j}}\bigg)^{s} \bigg(\frac{\partial}{\partial v_{j}}\bigg)^{L-s}  J_{j} \\ &=& J_{j} \sum_{} \prod_{i=0}^{s}\frac{v_{k_{i}}}{u_{j}v_{k_{i}} - u_{k_{i}}v_{j}} \prod_{i=0}^{L-s} \frac{-u_{l_{i}}}{u_{j}v_{l_{i}} - u_{l_{i}}v_{j}} \, ,
\end{eqnarray}
where the sum runs over all possibilities of choosing $s$ particles $\{k_{i}\}$ and $L{-}s$ particles $\{l_{i}\}$, all distinct. The Jastrow factors $J_{j}$ can be factored out of the determinant post-projection and their product is the square of the complete Jastrow factor, i.e., $\prod_{j}J_{j}=\prod_{i<k}(u_i v_k - u_k v_i)^2 \equiv \psi^{\rm L}_{2}$.

The above construction produces CFL states at general fillings. For the fermionic state at $\nu=1/4$, we choose to only project with two Jastrow factors, leaving the other two outside -- this is similar to the $\boldsymbol{\alpha}=(4,0)$ state on the torus. For the bosonic states at $\nu=1$ and $\nu=1/3$, we need to divide or multiply by a single Jastrow factor respectively~\cite{Chang05b, Liu20}, again mirroring the construction on the torus. It is known that the microscopic CFL wave function is not very sensitive to the number of Jastrows placed inside the LLL-projection operator~\cite{Balram15a, Balram16b}. Note that  $\nu=1$ is a special case as the hardcore constraint is not enforced properly, which can result in instability after projection. To fix this, we implicitly enforce a non-zero hardcore radius in our Monte Carlo simulation. 

\begin{figure}[tb]
    \centering    \includegraphics[width=\linewidth]{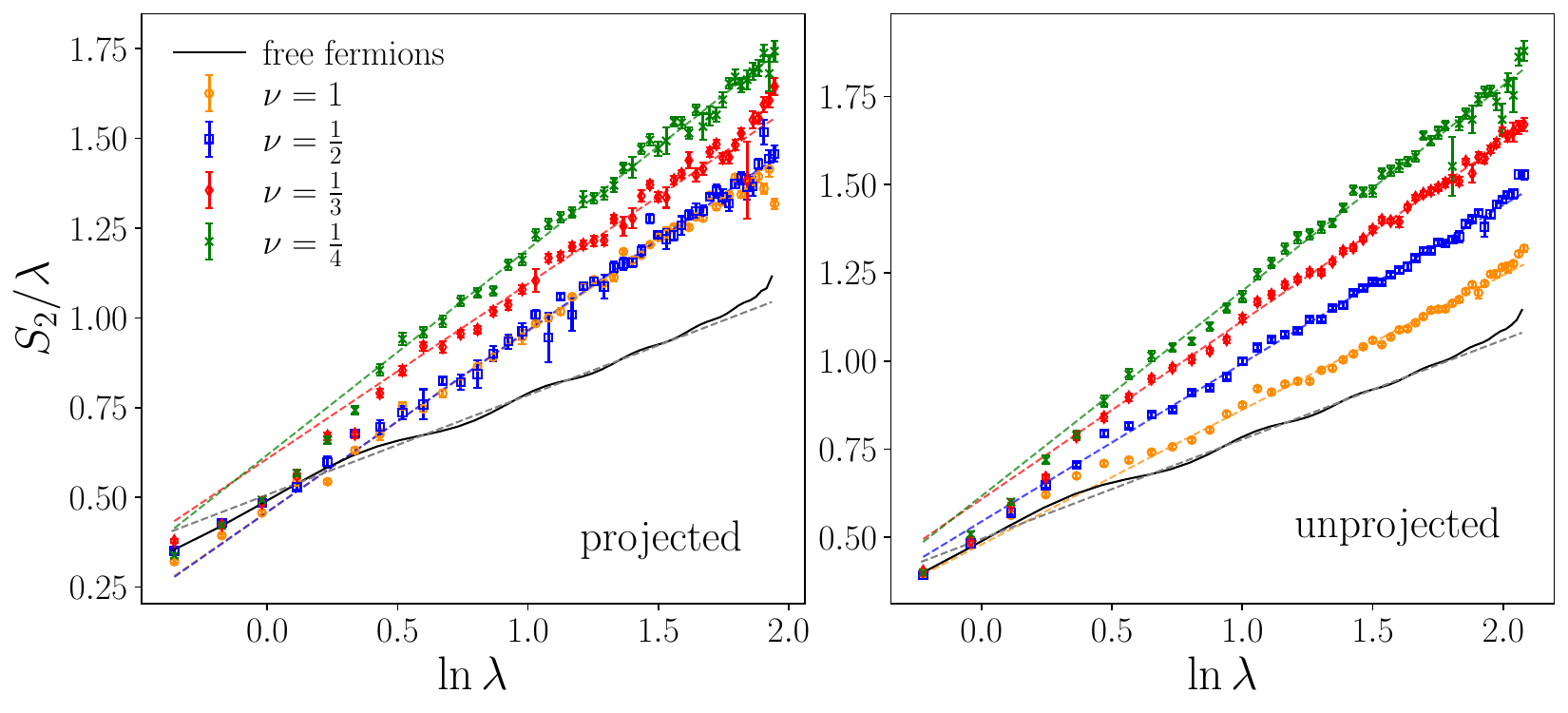}
    \caption{$S_2$ R\'enyi entropy of projected and unprojected CFL states on the sphere at fillings $\nu  = 1/m$, with $m=1,2,3,4$. The projected data is for the system size $N=49$, while the unprojected data is for $N=64$. The continuous black line is free-fermion data obtained by exact diagonalization. All filling factors show an enhanced slope compared to the Widom value, and $\nu=1$ shows an increase after projection, similar to the torus geometry. 
    } 
    \label{fig: entropy total all fillings sphere}
\end{figure}
Fig.~\ref{fig: entropy total all fillings sphere} demonstrates the similarity of CFL results on the sphere with those presented previously for the torus geometry. This consistency highlights the robustness of our results. Furthermore, for the sequence of system sizes $N=n^2$ we consider on the sphere, the CFL is a uniform angular momentum $L=0$ state, which removes some of the ambiguity associated with the definition of the Fermi momentum on the torus.

\section{Charge fluctuations}\label{sec:charge}

In systems with a U(1) symmetry, entanglement can be directly related to the fluctuations of particle number, $\hat N_A$, measured on a subsystem:
\begin{eqnarray}\label{eq:varN}
 \Delta^2 N_A \equiv \langle \hat N_A^2 \rangle - \langle \hat N_A\rangle^2 \;.
\end{eqnarray}
For example, in non-interacting 2D Fermi liquids, $S_2$ and $\Delta^2 N_A$ differ only by a multiplicative constant \cite{Calabrese2012,Tan20}. Similarly, in a NFL, charge fluctuations are expected to closely track entanglement entropy~\cite{SwingleSenthil2013}. 
Recently, the scaling of charge cumulants or ``disorder operators" was indeed shown to exhibit a logarithmic violation of the area law in NFLs at a quantum critical point~\cite{Jiang23}. Since $\Delta^2 N_A$ can be computed with a similar Monte Carlo method for the same system sizes as $S_2$, we can leverage the better analytical understanding of charge fluctuations to support our previous entanglement results.  

\begin{figure}[tb]
    \centering    
    \includegraphics[width=1\linewidth]{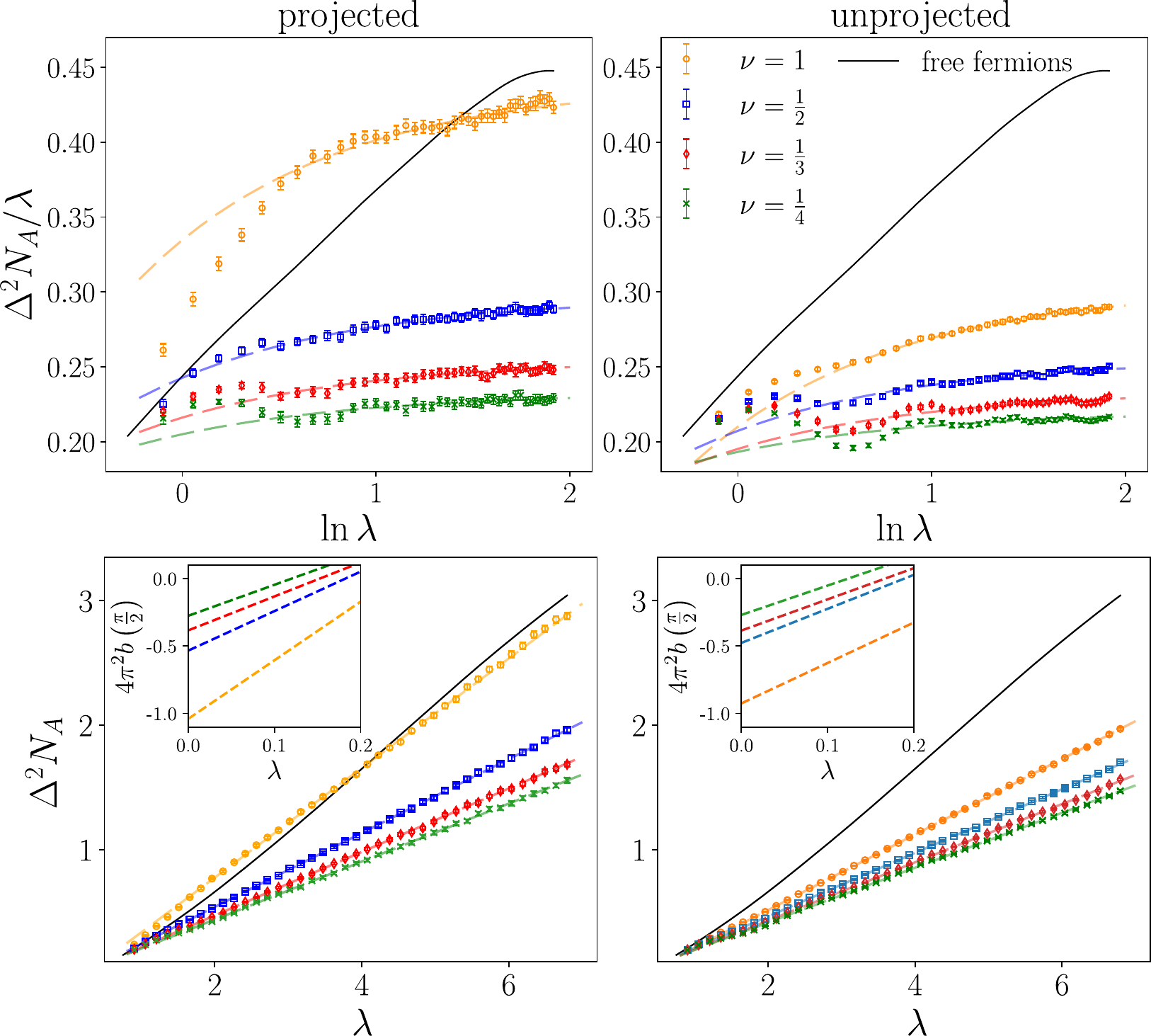}
    \caption{Charge fluctuations across the square-shaped bipartition for CFL states at different filling factors. In contrast to free fermions (solid black line), the charge fluctuations in CFL states are strongly suppressed and obey area-law scaling. The top row shows the fluctuations divided by $\lambda$ and plotted as a function of $\ln \lambda$, while the bottom row shows the raw value plotted on a linear $\lambda$ scale. Insets to the bottom plots show fits to the area law (colored dashed lines). LLL projection alters the area law coefficient but preserves the universal corner contribution. \Cref{tab: corner contributions} shows the extracted corner contributions, which are proportional to the filling factor.
    } 
    \label{fig: torus fluctuations}
\end{figure}

Remarkably, Fig.~\ref{fig: torus fluctuations} shows that $ \Delta^2 N_A$ in CFL states, with and without LLL projection, obeys an area law.  
This implies that charge fluctuations of the physical electrons forming the CFL are gapped~\cite{Wu2024bipartite, Cai2024disorder}, in stark contrast with ordinary Fermi liquids. While it is well-known that the $\nu{=}1/2$ CFL state does not exhibit a quantized Hall resistance~\cite{Jiang89}, transport experiments probe a gap to creating or removing fractionalized excitations. By contrast, $ \Delta^2 N_A$ probes the fluctuation of electrons in a subsystem and is therefore more similar to scanning tunneling spectroscopy experiments~\cite{Ashoori90, eisenstein92b, Xiaomeng2022, Farahi2023, Hu23}, which indeed observe a Coulomb gap at $\nu{=}1/2$. This is because the electrons in the CFL exhibit an effectively \emph{finite} correlation length leading to an exponentially suppressed tunneling density of states \cite{Barkeshli15, He93, Kim94, Pu2023}. The area-law for $ \Delta^2 N_A$ obeyed by the microscopic CFL wave function, to the best of our knowledge, has been demonstrated for the first time in this work.

In spatially isotropic systems, the static structure factor $s_{q}$ governs the behavior of $\Delta^2 N_A$~\cite{Berthiere2023}. For CFL states, 
the LLL-projected structure factor takes the form $\bar{s}_{q} \sim q^{3} \ln (1/q)$ in the long-wavelength limit~\cite{HalperinLeeRead, Son15}, with inter-Landau level gapped modes contributing an additional term $\sim q^{2}$ to $s_{q}$. The former enforces an area law scaling, while the latter adds a ``super-universal" constant term in the presence of sharp corners. The general form is 
\begin{equation}\label{eq: fluctuations}
    \Delta^{2}N_{A} = a \lambda - b(\theta) + \ldots, \;\;\;   b(\theta) = \frac{\nu}{4\pi^{2}}(1 + (\pi - \theta) \cot \theta) ,
\end{equation}
where $a$ is a non-universal coefficient, and $b(\theta)$ is a constant contribution due to a bipartition containing a single corner with an opening angle $\theta$~\cite{Wu2024bipartite}. In Fig.~\ref{fig: torus fluctuations}, for simplicity, we consider a square subsystem.

From the intercept of $\Delta^2 N_A$ vs. $\lambda$, we determined $b(\pi/2)$ in Table~\ref{tab: corner contributions}, which is in good agreement with Eq.~\eqref{eq: fluctuations}. Note that while projection places the wave functions of Eq.~\eqref{eq:WF} completely inside the LLL, we do not project the density operator, and therefore the associated gapped modes are not removed, hence both projected and unprojected wave functions exhibit approximately the same corner contribution. 

\begin{table}[!htb]
\centering
\begin{tabular}{|c|c|c|c|c|}
    \hline
    & $\nu {=} 1$ & $\nu{=}1/2$ & $\nu{=}1/3$ & $\nu{=}1/4$ \\
    \hline
    $4\pi^{2}b_\mathrm{theory}$ & 1 & 0.5 & 1/3 & 0.25 \\
    \hline
    $4\pi^{2}b_{\mathrm{proj}}$ & 1.04(9) & 0.53(4) & 0.38(6) & 0.28(6) \\
    \hline
    $4\pi^{2}b_{\mathrm{unproj}}$  & 0.93(3) & 0.48(4) & 0.39(4) & 0.27(5) \\
    \hline
\end{tabular}
\caption{Extracted corner contributions to charge fluctuations in different CFL states, with or without LLL projection. The system size is $N{=}37$ particles for all fillings.}
\label{tab: corner contributions}
\end{table}

\begin{figure*}[htb]
    \centering    
    \includegraphics[width=\linewidth]{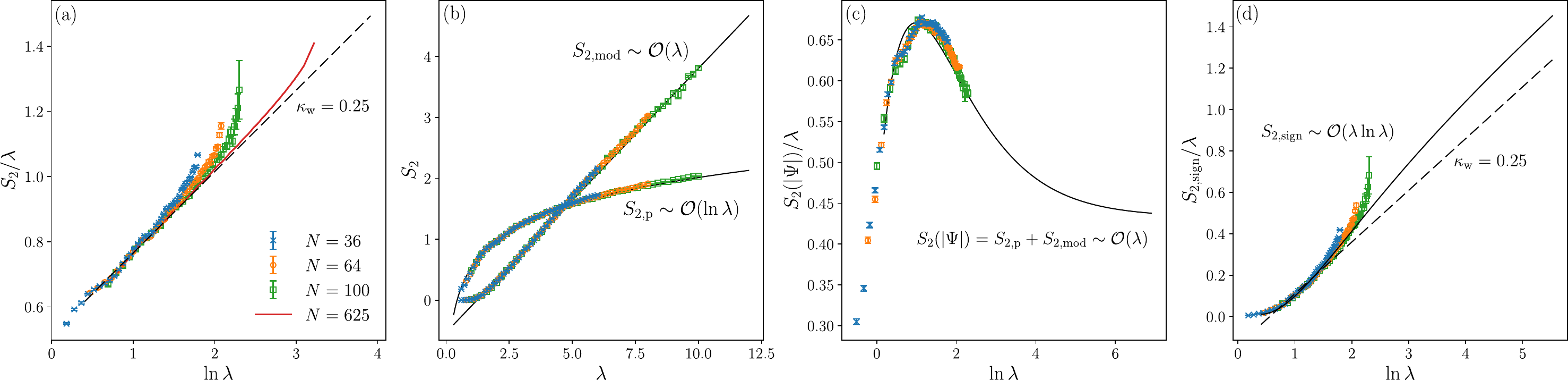}    
    \caption{Comparison of the full $S_2$ R\'enyi entropy and its SWAP decomposition, Eq.~\eqref{eq:modsign}, for free fermions on the sphere. 
    (a) Total $S_2$ entropy, computed via Monte Carlo (markers), while exact diagonalization result (solid line) is for a large system size $N=625$. All system sizes steadily approach the Widom slope $\kappa_{\mathrm{w}} {=} 0.25$ (dashed line). 
    (b): $S_{2,\mathrm{p}}$ and $S_{2,\mathrm{mod}}$ that sum up to the absolute value entropy $S_{2}(|\Psi|)$. $S_{2,\mathrm{p}}$ appears to scale logarithmically with the subsystem size, while $S_{2,\mathrm{mod}}$ follows an area-law. Black lines are $\lambda  \to \infty$ extrapolations. (c) The absolute value entropy $S_{2}(|\Psi|) = S_{2,p}+S_{2,\mathrm{mod}}$. The black line is the sum of the extrapolations in (b), which eventually settles to an area law for very large systems. However, unlike the individual decompositions in (b), their sum here appears far from the asymptotic scaling regime.  (d) The sign entropy $S_{2,\mathrm{sign}}$. The dashed line is the Widom slope, while the continuous black line is the Widom slope minus the extrapolation of $S_{2}(|\Psi|)$. This highlights the difficulty in extracting the Widom slope solely from the sign term: the continuous and dashed lines only become parallel for $\ln \lambda \gtrsim 4$, effectively requiring much larger system sizes to infer the correct scaling compared to the full entropy in (a).
    } 
    \label{fig: free fermions decomposition sphere}
\end{figure*}

\section{Modulus and sign structure}\label{sec:decomp}

While Fig.~\ref{fig: torus total entropy} strongly suggests that $\kappa$ exceeds the Widom value, it is important to understand where this enhancement is structurally encoded in a CFL wave function and whether or not it should be interpreted as a finite-size effect~\cite{Mishmash16}. An idea first put forward in Refs.~\cite{Zhang11, Zhang11PRB} is to decompose the total entropy in a way inspired by SWAP: 
\begin{equation}\label{eq:modsign}
    S_{2}(\Psi) =  S_{2,\mathrm{p}} + S_{2,\mathrm{mod}} + S_{2,\mathrm{sign}}.
\end{equation}
The first term represents the probability that
the two copies are ``swappable”, while the last two terms are the ``mod'' and ``sign'' contributions -- see Appendix~\ref{app: algorithm and decomp}. Adding the first two contributions together yields $S_{2}(|\Psi|)$ -- the entropy of the absolute value of the wave function, while $S_{2,\mathrm{sign}}$ stems from an interplay between the absolute value and sign structure of $\Psi$.

The hope behind the decomposition of Eq.~\eqref{eq:modsign} is that the sign structure of the wave function carries the leading contribution to the entanglement as $N{\to}\infty$. While there are examples of wave functions with provable large sign entanglement~\cite{Grover14, Kaplis17}, there are also instances where the mod entropy is not dominated by that of the sign \cite{Broecker16}. Thus, the scaling of different terms with $N$ in Eq.~\eqref{eq:modsign} is not rigorously understood in general. In fact, even for free fermions, the scalings of $S_{2}(|\Psi|)$ and  $S_{2,\mathrm{sign}}$ are not known analytically. Intuitively, the reason for this is that the decomposition in Eq.~\eqref{eq:modsign} replaces a tractable free-fermion problem with an effectively ``many-body'' object that no longer has a simple description in terms of correlation matrices~\cite{Chung01}. Moreover, $S_{2}(|\Psi|)$ and  $S_{2,\mathrm{sign}}$ carry non-trivial finite-size dependence due to the mixing of subleading [$\mathcal{O}(\lambda)$ and $\mathcal{O}(\ln \lambda)$] terms.
Our Monte Carlo analysis for free fermions presented below is consistent with  $S_{2,\mathrm{sign}}$ carrying the dominant $\lambda \ln \lambda$ dependence, however both $S_{2}(|\Psi|)$ and  $S_{2,\mathrm{sign}}$ suffer from much more pronounced finite-size effects compared to the total entropy $S_2$. Below we discuss in detail the entropy decompositions for free fermions and then apply the same considerations to the CFL case.

\subsection{Mod and sign entropy for free fermions}

As mentioned previously, the scaling of different contributions to the entropy in Eq.~\eqref{eq:modsign} is not rigorously known, even for free fermions where the full entropy has an established analytic expression.  Fig.~\ref{fig: free fermions decomposition sphere}(a) contrasts the full $S_2$ entropy for free fermions against its decompositions in Eq.~(\ref{eq:modsign}) obtained via Monte Carlo, Fig.~\ref{fig: free fermions decomposition sphere}(b)-(d). While the full entropy can be computed for large free-fermion systems containing thousands of particles, the evaluation of mod-sign entropy contributions, to the best of our knowledge, can only be done with Monte Carlo. This limits the accessible system sizes to $N \lesssim 100$, which is similar to the range for the CFL states.

Within the range $N \lesssim 100$, we observe the full entropy $S_2$ steadily converging to the Widom slope in Fig.~\ref{fig: free fermions decomposition sphere}(a). Figs.~\ref{fig: free fermions decomposition sphere}(b)-(c) illustrate the importance of considering the $\mathrm{p}$ and $\mathrm{mod}$ contributions to the entropy separately. They each appear to converge nicely to the respective scalings, $S_{2,p} \sim \mathcal{O}(\ln\lambda)$ and $S_{2,\mathrm{mod}} \sim \mathcal{O}(\lambda)$. However, if we add them together to make up $S_2(|\Psi|)$, Fig.~\ref{fig: free fermions decomposition sphere}(c), the latter converges non-monotonically when plotted as $S_2(|\Psi|)/\lambda$ vs $\ln\lambda$. If we estimate the slope of $S_2(|\Psi|)$ by adding the fits, we obtain the solid line in Fig.~\ref{fig: free fermions decomposition sphere}(c), which suggests that our system sizes are not large enough to see the expected area-law scaling $S_2(|\Psi|) \sim \lambda$  in the thermodynamic limit. This adversely impacts the scaling of the sign entropy in Fig.~\ref{fig: free fermions decomposition sphere}(d), which overshoots the Widom slope to compensate for the downturn in $S_2(|\Psi|)$. By plotting the Widom slope minus the extrapolation of $S_{2}(|\Psi|)$ using a solid line in Fig.~\ref{fig: free fermions decomposition sphere}(d), we see that it might only approach the Widom scaling in much larger system sizes ($\ln\lambda \gtrsim 4$). At the same time, the full entropy in Fig.~\ref{fig: free fermions decomposition sphere}(a) tracks the Widom scaling much more closely even in smaller system sizes.  As the range of system sizes in this example is comparable to the CFL case, the usefulness of mod and sign decompositions over the full entropy is questionable.

\subsection{Mod and sign entropy for CFL}

\begin{figure}[tb]
    \centering    \includegraphics[width=1\linewidth]{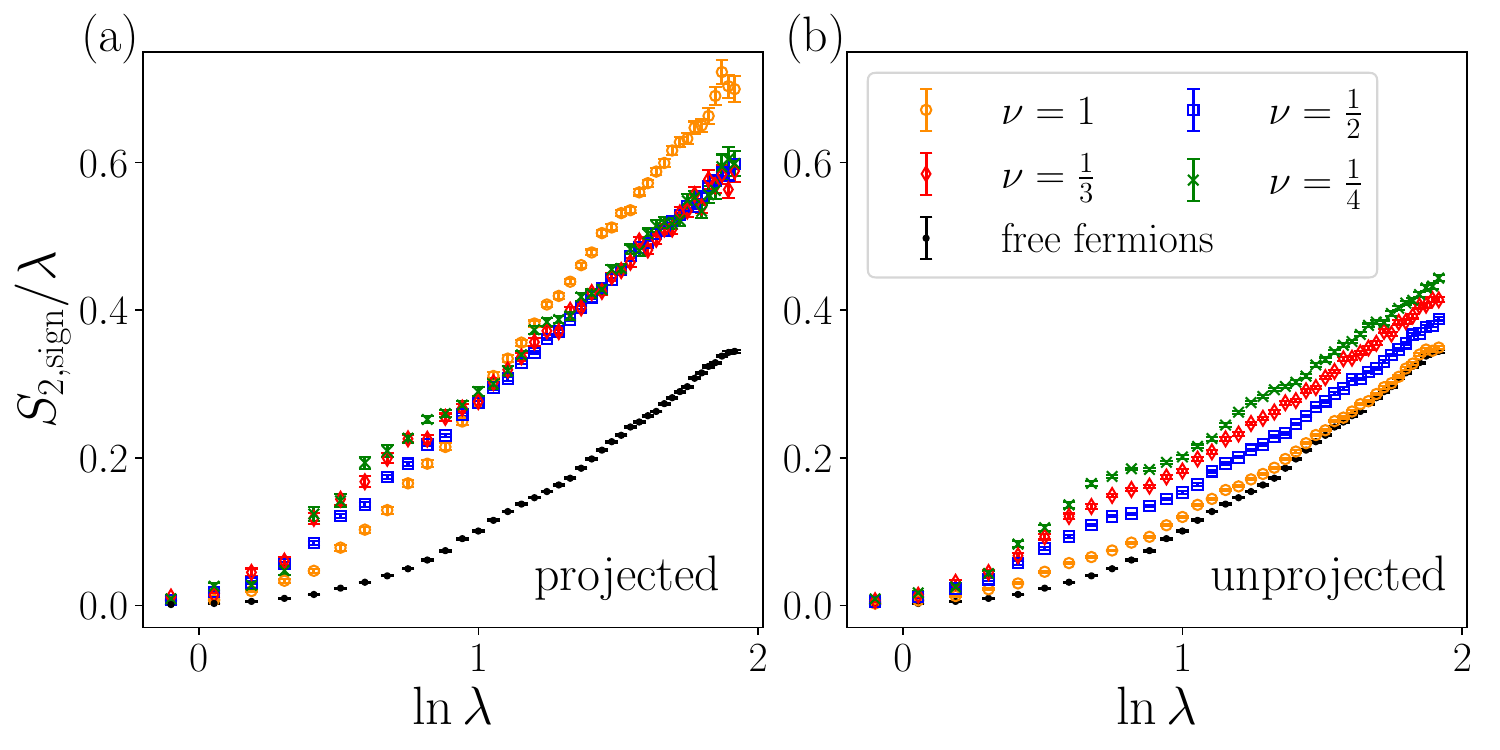}     \includegraphics[width=1\linewidth]{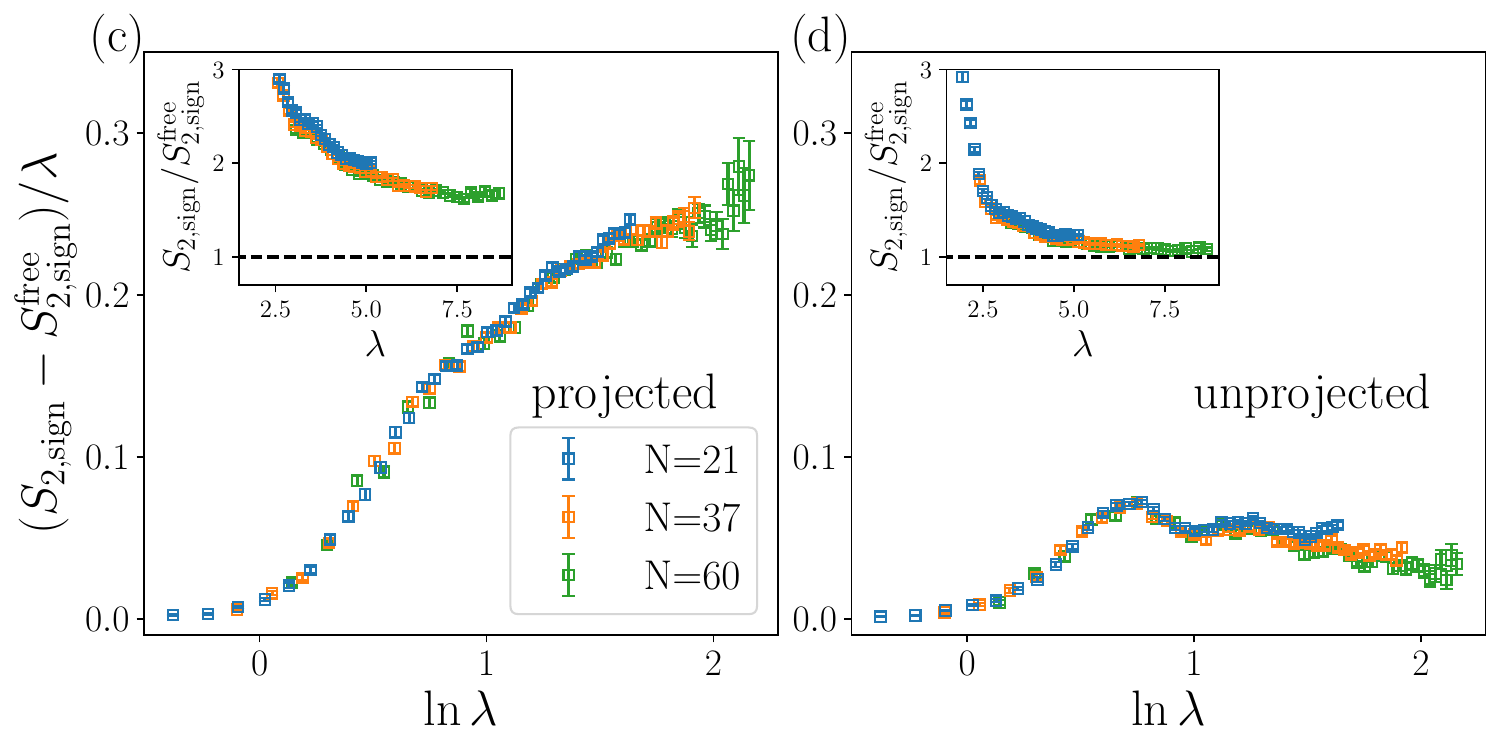}      
    \caption{Sign entropy for projected (a) and unprojected (b) CFL states for $N=37$ particles on a torus. For comparison, we also show the result for free fermions, with the same data plotted in (a) and (b). Unprojected CFL states behave similarly to free fermions, while projected wave functions show an enhanced slope. This is revealed by plotting the difference of $S_{2,\mathrm{sign}}$ for the $\nu=1/2$ CFL relative to free fermions in panels (c)-(d): the difference grows in the projected case (c), while it decays towards zero in the unprojected case (d). Insets in (c)-(d) show the ratio of CFL sign entropy to that of free fermions, which shows qualitatively similar behavior. 
    }
    \label{fig: torus sign entropy}
\end{figure}

Similar issues are observed in the sign entropy of projected and unprojected CFL states, Fig.~\ref{fig: torus sign entropy}(a)-(b). Unlike the full entropy, which is largely insensitive to the LLL projection, the sign entropies are significantly altered by it, although they still appear to capture the logarithmic violation of the area law. In particular, $S_{2,\mathrm{sign}}$ of the unprojected CFL states tracks the sign entropy of free fermions, which is revealed more clearly by plotting their difference in Fig.~\ref{fig: torus sign entropy}(d), while the ratio is shown in the inset. By contrast, within the same range of system sizes, the difference $S_{2,\mathrm{sign}}{-}S_{2,\mathrm{sign}}^\mathrm{free}$ \emph{increases} with $\lambda$ for the projected CFL states in Fig.~\ref{fig: torus sign entropy}(c), consistent with an enhancement. 

As the physical content and scaling properties of the sign entropy are unclear, one should not view the results in Fig.~\ref{fig: torus sign entropy} as more compelling than the full entropy in Fig.~\ref{fig: torus total entropy}. Nevertheless, if one assumes that $S_{2,\mathrm{sign}}$ may carry the full $\lambda \ln \lambda$ dependence, it is interesting to ponder two scenarios that would make Figs.~\ref{fig: torus total entropy} and~\ref{fig: torus sign entropy} consistent.  The first possibility is that there may be an intrinsic difference between projected and unprojected CFL states, such that only the former exhibits an enhanced entanglement growth. While this is counterintuitive, LLL projection could alter the coefficient of entanglement scaling, unlike other universal topological properties. 
The second option, proposed by Ref.~\cite{Mishmash16}, is to conclude there is no enhancement and place trust in Figs.~\ref{fig: torus sign entropy}(b)-(d) above all others. This would require the lines in  Figs.~\ref{fig: torus sign entropy}(a)-(c) to ultimately undergo a downturn, with a corresponding reduction of slope in Fig.~\ref{fig: torus total entropy}. Based on the existing data, we estimate that observing this would require system sizes $N{\gtrsim}300$, which are significantly beyond current computational capability.

\section{Conclusions and discussion}\label{sec:conclusion}

We have evaluated the scaling of $S_2$ R\'enyi entropy of CFL states at different fillings $\nu{=}1/m$ and showed that it obeys the Widom scaling as in ordinary Fermi liquids but with an enhanced prefactor. We have found an enhancement in both the sphere and torus geometry; moreover, for $m{\geq}2$, the enhancement is essentially unaffected by the LLL projection. We have corroborated these conclusions by demonstrating that the wave functions in Eq.~\eqref{eq:WF} encode the expected area law for charge fluctuations with a universal corner contribution. Moreover, we pointed out the difficulties of interpreting the entropy scalings of the modulus and sign parts of the wave function, arguing that the total entropy is a more robust quantity in the range of numerically accessible system sizes.

Our results for $m{=}2$ are in agreement with Ref.~\cite{Shao15}, while the discrepancy with Ref.~\cite{Mishmash16} is, at least partly, due to the LLL projection which was not explicitly enforced on the lattice. 
In fact, our unprojected results in Fig.~\ref{fig: torus sign entropy}(b) are qualitatively similar to Ref.~\cite{Mishmash16}; by contrast, the LLL-projected results in Fig.~\ref{fig: torus total entropy} show a visible departure from Ref.~\cite{Mishmash16}, e.g., the entropy of the bosonic $\nu{=}1$ CFL is similar to that of the fermionic $\nu{=}1/2$ CFL state after LLL projection, unlike their lattice analogs. 
These differences point to a subtle role of LLL projection, precluding direct comparisons between continuum and lattice versions of the CFL state. This could potentially imply that the entanglement scaling may depend on the microscopic details of flux attachment (see Appendix~\ref{app:fluxattached}). Unfortunately, as we explained, settling this question numerically requires access to system sizes several times larger than the current computational facility. 

One remaining enigma is how to reconcile our results with the iDMRG study of the $\nu{=}1/2$ CFL state on an infinite cylinder, where no enhancement to the Widom formula was found~\cite{Geraedts16}. Two important differences in the setup of Ref.~\cite{Geraedts16} are the quasi-1D geometry, which can impact the behavior of the 2D structure factor~\cite{Kumar22}, and the study of the ground state of the Coulomb interaction rather than a variational wave function. Properties of the CFL for long-range interactions are known to be more similar to those of an ordinary Fermi liquid, while the differences are amplified by shorter-range interactions~\cite{HalperinLeeRead}. While our wave function [Eq.~\eqref{eq:WF}] is not an exact ground state of any known Hamiltonian, it would be interesting to repeat the analysis of Ref.~\cite{Geraedts16} and check the impact of short-range interaction on $\kappa$. 

Ultimately, the enhanced entanglement scaling in CFL states should be understood analytically. We expect that further progress could be made using multi-dimensional bosonization~\cite{Ding12, Cai2024disorder}, or by finding other observables whose fluctuations violate the area law, as recently achieved for the gapless Mott insulator with a spinon Fermi surface \cite{Jiang23, Wu2024bipartite}.  This understanding may shed light on other pertinent questions, such as whether the Chern band projection similarly enhances the entanglement of a CFL in zero magnetic field~\cite{Dong23, Goldman23}, or whether the latter also depends on band geometry and Chern number.

\begin{acknowledgments}	

We are grateful to Luca Delacr\'etaz, Benoit Estienne, Nicolas Regnault, Xiao-Chuan Wu, and Kun Yang for useful discussions, and to Ryan Mishmash for generously sharing his lattice Monte Carlo code with us and for helpful comments on our manuscript. C.V., S.P., and Z.P.  acknowledge support by the Leverhulme Trust Research Leadership Award RL-2019-015 and by EPSRC grants EP/R020612/1, EP/W026848/1. Statement of compliance with EPSRC policy framework on research data: This publication is theoretical work that does not require supporting research data. A.C.B. acknowledges support by the Royal Society International Exchanges Grant IES$\backslash$R2$\backslash$202052 and the Science and Engineering Research Board (SERB) of the Department of Science and Technology (DST) for financial support through the Mathematical Research Impact Centric Support (MATRICS) Grant No. MTR/2023/000002. This research was supported in part by grant NSF PHY-2309135 to the Kavli Institute for Theoretical Physics (KITP). Computational portions of this research work were carried out on ARC3 and ARC4, part of the High-Performance Computing facilities at the University of Leeds, UK. 
\end{acknowledgments}

\appendix

\begin{figure}[bt]
    \centering    
    \includegraphics[width=\linewidth]{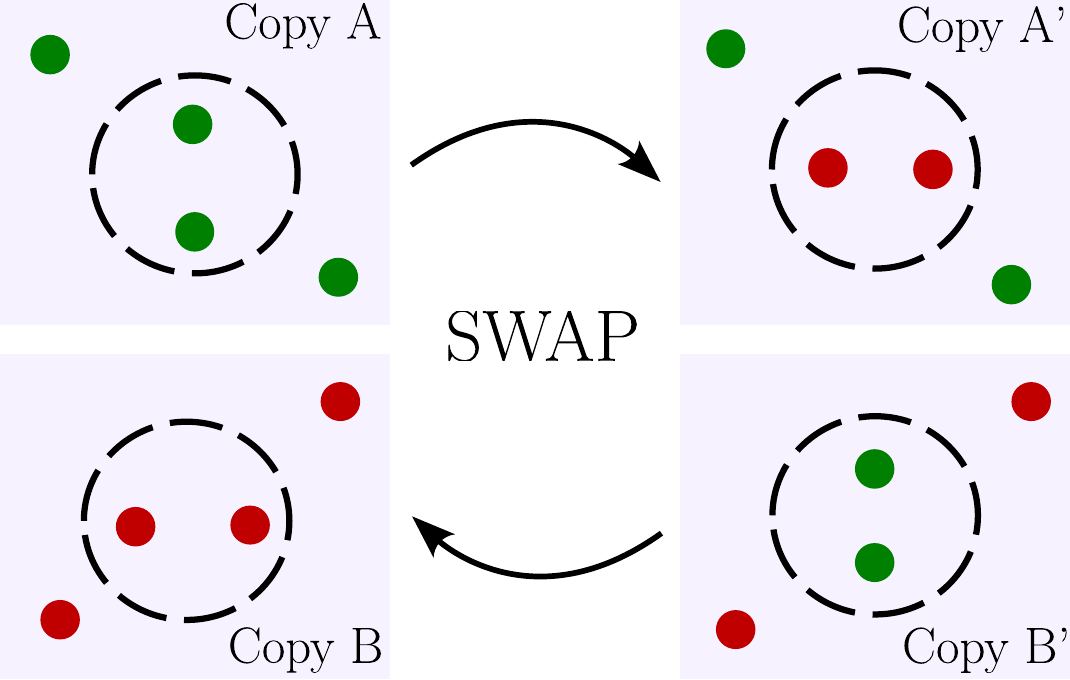}
    \caption{Illustration of the SWAP process. We have two copies of the full system, whose particles are labeled with different colors, and each copy is divided into two regions separated by the dashed circle. Through the SWAP process, we interchange the particle coordinates inside the circle between the two copies while keeping the outside coordinates unchanged.
    } 
    \label{swap}
\end{figure}

\section{SWAP algorithm and entropy decomposition} \label{app: algorithm and decomp}

Following Refs.~\cite{Hastings10, Zhang11}, in our Monte Carlo evaluations of $S_2$ R\'enyi entropy we replace the partial trace with an expectation value $ \langle \mathrm{SWAP}_{A} \rangle$ of the SWAP operator between two copies of the system, as illustrated in Fig.~\ref{swap}. This expectation value can be naturally broken down into a product of three terms:
\begin{equation}\label{eq: swap decomposition}
    \langle \mathrm{SWAP}_{A} \rangle = P_{\mathrm{SWAP}} \, \langle \mathrm{SWAP}_{\mathrm{mod}} \rangle \langle \mathrm{SWAP}_{\mathrm{sign}} \rangle \, .
\end{equation}
The first term, $P_{\mathrm{SWAP}}$, represents the probability that the two copies are ``swappable'', for which we must have the same number of particles $N_{1}=N_{2}$:
\begin{align}\label{eq:Sp}
    P_{\mathrm{SWAP}} &= \frac{\int \mathrm{d}\boldsymbol{\alpha}\mathrm{d}\boldsymbol{\beta}\, |\Psi(\alpha_{1},\beta_{1})|^{2} |\Psi(\alpha_{2},\beta_{2})|^{2} \delta_{N_{1}, N_{2}}}{\int \mathrm{d}\boldsymbol{\alpha}\mathrm{d}\boldsymbol{\beta}\, |\Psi(\alpha_{1},\beta_{1})|^{2} |\Psi(\alpha_{2},\beta_{2})|^{2}} \, ,
\end{align}
where $\boldsymbol{\alpha} = (\alpha_{1},\alpha_{2})$ and $\boldsymbol{\beta} = (\beta_{1},\beta_{2})$ denote the sets of coordinates for the particles inside or outside the bipartition contour, respectively. The resulting entropy $S_{2,\mathrm{p}}$ is the higher R\'enyi counterpart of the von Neumann number entropy \cite{Lukin19}, which can be generalized as $S_{\alpha,\mathrm{p}} = \frac{1}{1-\alpha} \ln \sum_{N_{A}} p(N_{A})^{\alpha}$, where $p(N_{A})$ is the probability of having $N_A$ particles in the defined subregion.

The last two terms in Eq.~(\ref{eq: swap decomposition}) are the ``mod'' and ``sign'' contributions, which are evaluated only in the subspace of swappable copies ($N_{1}=N_{2}$), 
\begin{widetext}
\begin{align}
\langle \mathrm{SWAP}_{\mathrm{mod}} \rangle
    &= \frac{ \int \mathrm{d}\boldsymbol{\alpha}\mathrm{d}\boldsymbol{\beta}  \, \delta_{N_{1},N_{2}} |\psi(\alpha_{1},\beta_{1})|^{2}|\psi(\alpha_{2},\beta_{2})| ^{2} \left| \frac{\psi(\alpha_{2},\beta_{1})\psi(\alpha_{1},\beta_{2})}{\psi(\alpha_{1},\beta_{1})\psi(\alpha_{2},\beta_{2})} \right|} {\int \mathrm{d}\boldsymbol{\alpha}\mathrm{d}\boldsymbol{\beta}\,  \delta_{N_{1},N_{2}} |\psi(\alpha_{1},\beta_{1})|^{2}|\psi(\alpha_{2},\beta_{2})| ^{2}}, \\
    \langle \mathrm{SWAP}_{\mathrm{sign}} \rangle
    &= \frac{ \int \mathrm{d}\boldsymbol{\alpha}\mathrm{d}\boldsymbol{\beta}\, \delta_{N_{1},N_{2}} |\psi(\alpha_{2},\beta_{1})\psi(\alpha_{1},\beta_{2}) \psi(\alpha_{1},\beta_{1})\psi(\alpha_{2},\beta_{2})| \,     e^{i\theta(\boldsymbol{\alpha}, \boldsymbol{\beta})}} {\int \mathrm{d}\boldsymbol{\alpha}\mathrm{d}\boldsymbol{\beta}\, \delta_{N_{1},N_{2}} |\psi(\alpha_{2},\beta_{1})\psi(\alpha_{1},\beta_{2}) \psi(\alpha_{1},\beta_{1})\psi(\alpha_{2},\beta_{2})|}\; .
\end{align}
\end{widetext}
In the sign contribution, the angle $\theta(\boldsymbol{\alpha},\boldsymbol{\beta})$ is defined as:
\begin{equation}
    \exp i \theta(\boldsymbol{\alpha},\boldsymbol{\beta}) = \frac{\psi(\alpha_{1},\beta_{2})^{*}\psi(\alpha_{2},\beta_{1})^{*} \psi(\alpha_{1},\beta_{1})\psi(\alpha_{2},\beta_{2})}{|\psi(\alpha_{1},\beta_{2})\psi(\alpha_{2},\beta_{1})\psi(\alpha_{1},\beta_{1})\psi(\alpha_{2},\beta_{2})|} \; .
\end{equation}
The contributions $P_{\mathrm{SWAP}}$, $\langle \mathrm{SWAP}_{\mathrm{mod}} \rangle$, and $\langle \mathrm{SWAP}_{\mathrm{sign}} \rangle$ yield separate contributions to the total R\'enyi entropy,
\begin{eqnarray}\label{eq:swapdecomp}
    S_2 = S_{2,\mathrm{p}} + S_{2,\mathrm{mod}} + S_{2,\mathrm{sign}}.
\end{eqnarray}
In previous works, such as Ref.~\cite{Mishmash16}, the first two contributions to the entropy were lumped together, because $S_{2,\mathrm{p}} + S_{2,\mathrm{mod}} \equiv S_2(|\Psi|)$ can be interpreted as the entropy of the absolute value of the wave function $\Psi$. While this may be useful in certain cases~\cite{Zhang11PRB}, our data below suggests this is not helpful for gapless CFL states since $S_{2,\mathrm{p}}$ and  $S_{2,\mathrm{mod}}$ exhibit different scaling with $\lambda$. Hence, we study individually the three distinct contributions to $S_2$.  

\section{Overlap integrals}\label{app:overlaps}

To evaluate $S_2$ R\'enyi entropy for free fermions in a finite-size system, we use Eq.~\eqref{overlap} which requires knowledge of the overlap integrals on the subsystem, $\mathcal{A}_{mn}$. 
In the torus geometry with a circular subregion $A$, the overlap matrix elements evaluate to: 
\begin{align}
    \notag \mathcal{A}_{mn}^{\mathrm{torus}} &= \int_A \mathrm{d}\mathbf{r}\,\phi_m^*(\mathbf{r})\phi_n(\mathbf{r}) = \frac{1}{L^{2}} \int_A \mathrm{d}\mathbf{r}\,e^{i(\mathbf{k}_{n}-\mathbf{k}_{m})\cdot \mathbf{r}} \\
    &= \frac{2\pi} {L^{2}} \frac{r_{A}}{|\mathbf{k}_{n}-\mathbf{k}_{m}|} J_{1} (|\mathbf{k}_{n}-\mathbf{k}_{m}| r_{A}) \, ,
\end{align}
where $J_{1}$ is the Bessel function of the first kind. On the sphere, for a spherical cap subregion, we obtain
\begin{align}
    \notag \mathcal{A}_{mn}^{\mathrm{sphere}} &= \int_{A} \mathrm{d}\Omega\, Y^{*}_{L_{m}M_{m}} (\Omega) Y_{L_{n}M_{n}} (\Omega) \\
    \notag &= 2 \pi \mathcal{N}_{L_{m}M_{m}}\mathcal{N}_{L_{n}M_{n}} \delta_{M_{m},M_{n}} \\
    &\times \int_{\cos \theta_{A}}^{ 1} \mathrm{d}x \, P_{L_{m}}^{M_{m}} (x) P_{L_{n}}^{M_{n}} (x) \, ,
\end{align}
where $\mathcal{N}_{LM} = \sqrt{\frac{2L+1}{4\pi}\frac{(L+M)!}{(L-M)!}}$ is a normalization constant and the associated Legendre polynomial integrals can be computed recursively.

\section{Comparison of R\'enyi entropies obtained  with different Jain-Kamilla projections}\label{app:JK}

\begin{figure}[bth]
    \centering    
    \includegraphics[width=\linewidth]{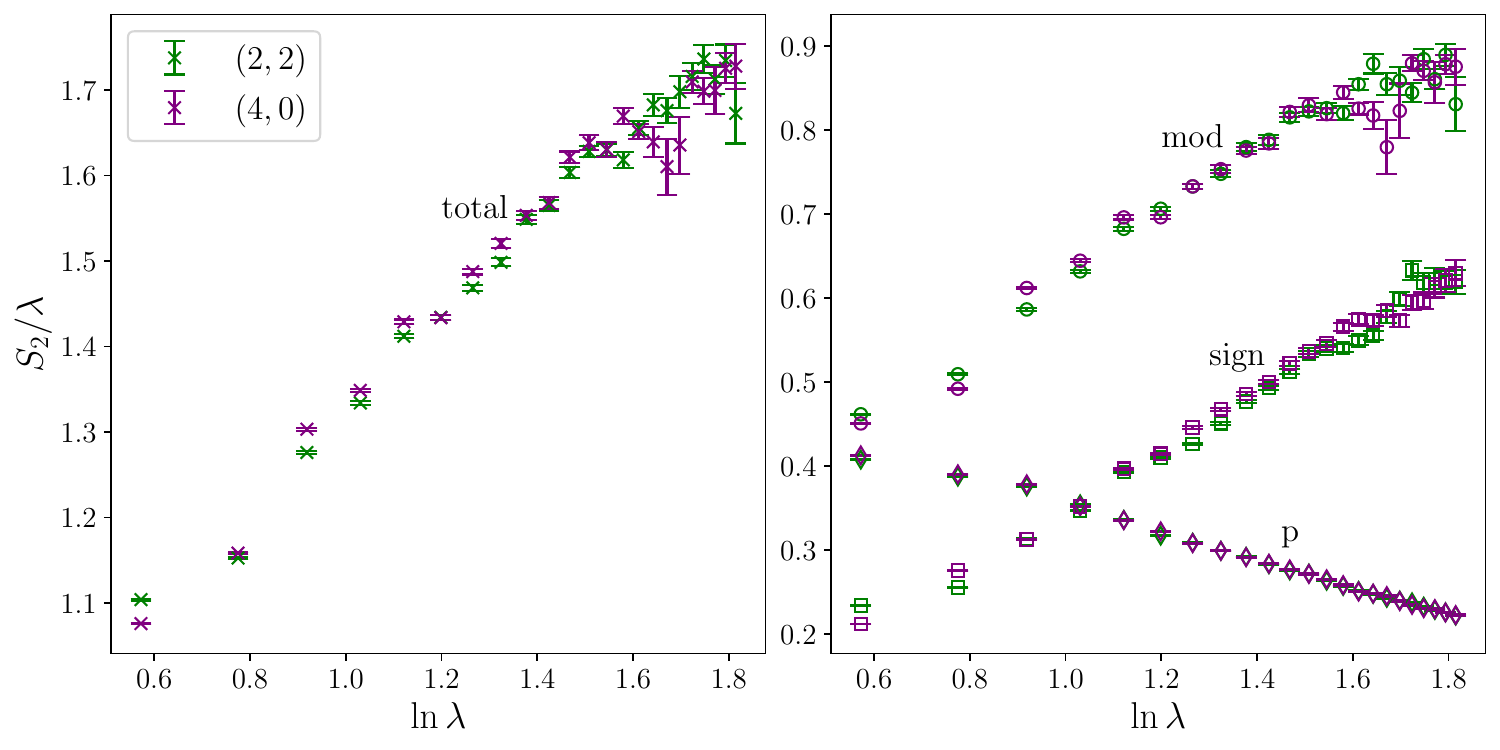}
    \caption{The $S_{2}$ R\'enyi entropy (left panel) and its decomposition (right panel) for the two choices of JK projection coefficients,  $\boldsymbol{\alpha}=(2,2)$ and  $\boldsymbol{\alpha}=(4,0)$, at filling factor $\nu=1/4$. The differences are seen to be minimal, the slope in particular agrees within the error bars, suggesting that details of the JK projection scheme are unimportant in determining the entanglement scaling. All data is for the system size of $N=37$ electrons on the torus.
    } 
    \label{fig: v=4 comparison}
\end{figure}

To verify the consistency of our results for different choices of $\boldsymbol{\alpha} $, Fig.~\ref{fig: v=4 comparison} shows the $S_2$ R\'enyi entropy scaling is nearly unchanged for two choices  $\boldsymbol{\alpha}=(2,2)$ and  $\boldsymbol{\alpha}=(4,0)$ at $\nu=1/4$. Moreover, even the individual contributions to the entropy, $S_{2,p}$, $S_{2,\mathrm{mod}}$ and $S_{2,\mathrm{sign}}$, discussed in Appendix~\ref{app: algorithm and decomp}, are essentially the same for the two $\boldsymbol{\alpha}$ choices. This shows that $S_2$ R\'enyi entropy is insensitive to the details of JK projection coefficients, unlike some other quantities such as the Hall viscosity in Ref.~\cite{Pu20}. 

\section{Slope extrapolation and shape dependence of the entanglement scaling}\label{app:extrapolation}

\begin{figure*}[htb]
    \centering    
    \includegraphics[width=0.34\linewidth]{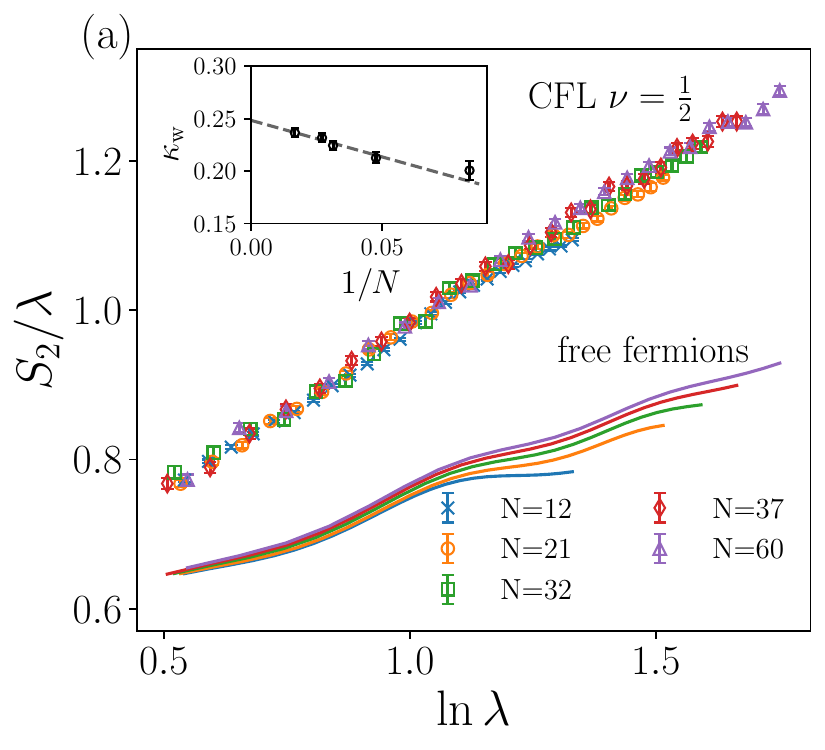}
    \includegraphics[width=0.62\linewidth]{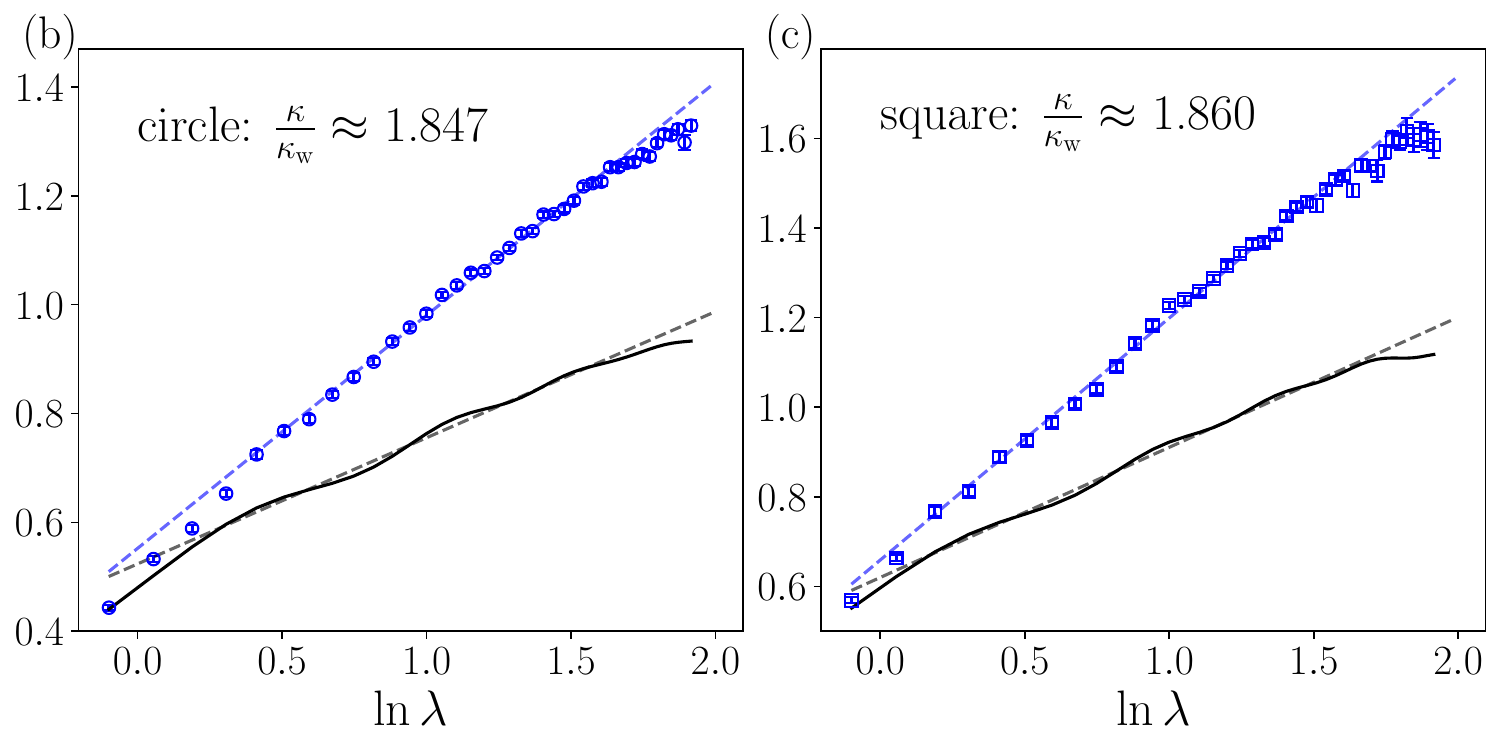}
    \caption{Slope extrapolation and shape dependence of the entanglement scaling. (a): We extract the $S_2$ slope from a subset of data points in the $\lambda$ range shown here for various system sizes indicated in the legend. Solid lines are free-fermion data, while markers denote the projected CFL data at $\nu{=}1/2$. The range of $\lambda$ values is chosen with free fermions as a benchmark, such that the correct Widom slope is recovered in the thermodynamic limit (inset). (b)-(c): The dependence of the entanglement scaling on the shape of the subregion, i.e., circle in (b) or square in (c). $S_{2}$ for free fermions (solid lines) and the projected CFL state at $\nu=1/2$ (markers), for system size $N{=}37$. The ratio of slopes is approximately independent of subregion geometry.
    } 
    \label{fig: SM extrapolation}
\end{figure*}

Here we present additional data on the slope extrapolations used in Fig.~\ref{fig: torus total entropy}(c) of the main text. Fig.~\ref{fig: SM extrapolation}(a) shows the data points used at system sizes $N\in \{12,21,32,37,60\}$ for the projected CFL wave function at $\nu{=}1/2$. We fix the smallest value to $\ln \lambda \approx 0.5$ regardless of system size, while the endpoint increases with $N$ such that $\ln \lambda _{\mathrm{end}} \in \{1.33,1.51,1.59,1.66,1.75\}$. These values are chosen to maximize the range of data points used at each system size, while still correctly extrapolating the correct Widom coefficient in the thermodynamic limit -- see inset of Fig.~\ref{fig: SM extrapolation}(a). 

Fig.~\ref{fig: SM extrapolation}(b) and (c) show a comparison between systems with subregions of circular and square shape, respectively. We find the ratio of the slopes to be approximately independent of the subregion geometry. This indicates that the CFL entanglement scaling still encodes the geometric information in the Widom formula, with the violation limited to a numerical overall prefactor.

\section{The Widom formula on the sphere}\label{app:widomsphere}

The intuitive derivation of the Widom formula given in Ref.~\cite{Swingle10} requires a few modifications in the spherical geometry. Suppose we have $N=n^2$ particles filling up $n$ angular momentum shells. The $2n{-}1$ states of the last filled shell (the equivalent of the Fermi surface in the thermodynamic limit) are the ones that will contribute to the entanglement entropy. Therefore, one needs to replace the integral over the Fermi surface in the Widom formula,
\begin{equation} \label{eq:widomSMtorus}
\kappa_{\mathrm{w}}^{(\alpha)} =  \frac{c_{\mathrm{eff}}(1+\alpha)}{12\alpha} \frac{1}{2}  \int_{\partial \Omega} \mathrm{d}S_{x} \int_{\partial \Gamma} \mathrm{d}S_{k} \, \frac{|\hat{\mathbf{n}}_{x} \cdot \hat{\mathbf{n}}_{k}|}{2\pi},
\end{equation}
with a discrete sum over these modes. 
Here, $c_\mathrm{eff}$ denotes the effective central charge of a (1+1)-
dimensional chiral relativistic fermion, the vectors $\hat{\mathbf{n}}_{x}$ and $\hat{\mathbf{n}}_{k}$ are unit normals for real space boundary and the Fermi surface, respectively, and $\alpha$ denotes the R\'enyi index (we will mainly focus on $\alpha=2$).

The flux factor $|\hat{n}_{x}\cdot \hat{n}_{k}|/2\pi$ needs to be replaced by $1/2\pi$, accounting for the rotational symmetry of the cap around the axis, and the double counting factor of $1/2$ becomes unnecessary. Then, the R\'enyi entropy for a spherical cap of angle $\theta$ is:
\begin{eqnarray} \label{eq:widomSMsphere}
\notag S_{\alpha} &=&  \frac{c_{\mathrm{eff}}(1+\alpha)}{12\alpha}  \ln (R \sin \theta) \sum_{i=1}^{2n-1} \int_{\partial \Omega} \frac{\mathrm{d}S_{x}}{2\pi} \\
&=& \frac{c_{\mathrm{eff}}(1+\alpha)}{12\alpha} (2\sqrt{N}-1) \sin \theta \ln (R \sin \theta).
\end{eqnarray}
To enable a comparison between this and CFL states, we need to control the particle density through the flux. This, in turn, affects the definition of the radius $R$. For the CFL, the radius of the sphere is $R=\sqrt{Q}$ and the maximum occupied effective angular momentum is $L_\mathrm{max}=\sqrt{N}$. Setting $\alpha=2$,  the radius of the cap to $R\sin \theta=\sqrt{Q} \sin \theta$, and the effective Fermi wave number $k_{F}{=}L_\mathrm{max}/R=\sqrt{N/Q}$, we express the entropy in terms of the dimensionless length $\lambda_{s} {=} k_{F} \sqrt{Q} \sin \theta$ : 
\begin{equation}
    S_{2} = \left( \frac{1}{4} - \frac{1}{8\sqrt{N}} \right) \lambda_{s} \ln \lambda_{s}\approx \frac{1}{4} \, \lambda_{s} \ln \lambda_{s},
\end{equation}
in agreement with the torus in the large-$N$ limit. Note that our derivation here does not capture finite-size effects related to the curvature of the sphere, which diminishes in the limit $N\to\infty$.

\section{Flux-attached wave functions}\label{app:fluxattached}

In the main text, we have mentioned the possibility of the Widom formula losing its ``universal" geometric character in NFLs, such that the coefficient of the leading term depends on microscopic details of flux attachment. To further explore this, we calculate the R\'enyi entropy and its decomposition in ``flux-attached" wave functions, where the Jastrow factors are replaced by their phases:
\begin{equation}\label{eq:WF_fluxattached}
\Psi_{m}^\mathrm{CFL,flux} = \mathrm{Det} \big[ 
\chi_{n} (z_{j}) \big] 
{\prod}_{i{<}j} \left( \frac{(z_i{-}z_j)}{|z_{i}-z_{j}|}\right)^m e^{-\frac{1}{4}\sum_k |z_k|^2}.
\end{equation}
For such a wave function, the mod entropy is identical to that of free fermions, placing any potential enhancement solely in the sign structure of the wave function.

\begin{figure}[tbh]
    \centering    
    \includegraphics[width=\linewidth]{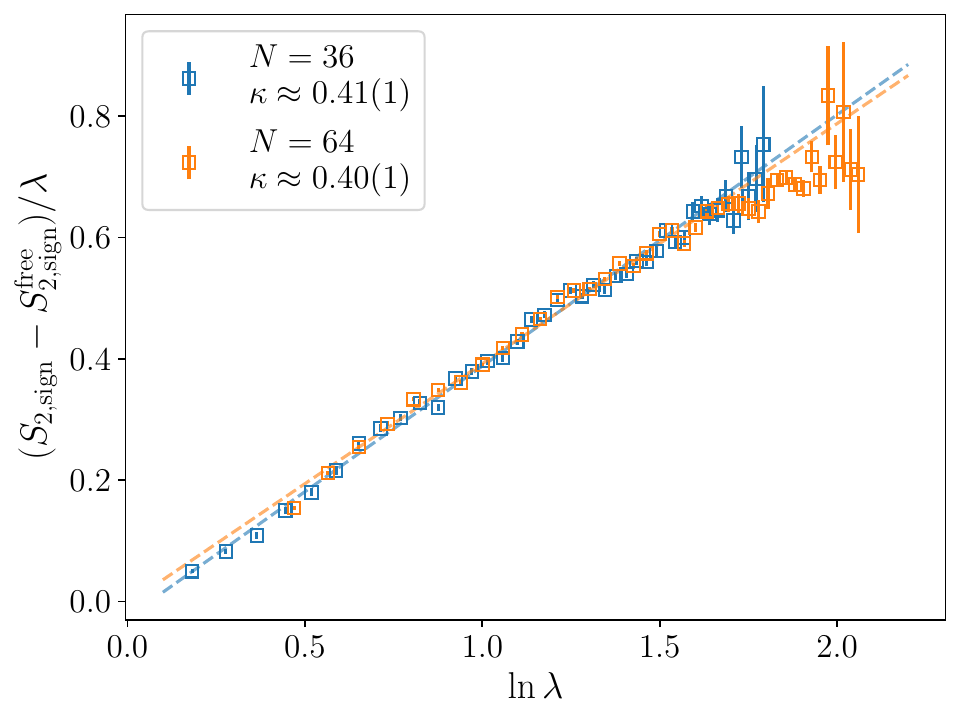} 
    \caption{Difference in sign entropy $S_{2,\mathrm{sign}} - S_{2,\mathrm{sign}}^{\mathrm{free}}$ between the flux-attached CFL at filling $\nu{=}1/2$ and free fermions, on the sphere. We notice that the entanglement scaling of the sign entropy is considerably enhanced.}
    \label{fig: flux attachment}
\end{figure}

\Cref{fig: flux attachment} shows the results for $S_{2,\mathrm{sign}}$ of the state in Eq.~\eqref{eq:WF_fluxattached}, with $m=2$ in the spherical geometry. We see that the sign entropy is significantly enhanced, even surpassing the standard vortex-attached wave functions discussed in the main text, which is in good agreement with the lattice realization of the flux-attached wave function \cite{MishmashPrivateComm}. Unfortunately, we are unable to efficiently project this wave function as a significant proportion of it resides outside the lowest Landau level.

\bibliography{biblio_fqhe}

\end{document}